\documentclass[a4paper,UKenglish,cleveref, autoref, thm-restate]{oasics-v2021}

\usepackage[nolist]{acronym}
\usepackage{amsmath}
\usepackage{amsfonts}
\usepackage{algorithm}
\usepackage{algorithmic}
\usepackage{tabularray}

\begin{acronym}
    \acro{GPUs}{Graphics Processing Units}
    \acro{TPUs}{Tensor Processing Units}
    \acro{NPUs}{Neural Processing Units}
    \acro{CPUs}{Central Processing Units}
    \acro{FPGAs}{Field-Programmable Gate Arrays}
    \acro{SWaP-C}{Size, Weight, Power, and Cost}
    \acro{MCSs}{Mixed Criticality Systems}
    \acro{SPH}{Static Partitioning Hypervisors}
    \acro{LLC}{Last-Level Cache}
    \acro{WCET}{Worst-Case Execution Time}
    \acro{VMs}{Virtual Machines}
    \acro{MBR}{Memory Bandwidth Reser}
\end{acronym}

\title{SP-IMPact: A Framework for Static Partitioning Interference Mitigation and Performance Analysis}

\titlerunning{SP-IMPact}

\author{Diogo {Costa}}{Centro ALGORITMI / LASI, Universidade do Minho, Portugal}{diogocostaes21@gmail.com}{https://orcid.org/0000-0002-0725-5861}{Supported by FCT grant 2022.13378.BD}

\author{Gonçalo Moreira}{Centro ALGORITMI / LASI, Universidade do Minho, Portugal}{pg53841@alunos.uminho.pt}{https://orcid.org/0009-0009-9883-1403}{}

\author{Afonso Oliveira}{Centro ALGORITMI / LASI, Universidade do Minho, Portugal}{pg53599@alunos.uminho.pt}{https://orcid.org/0009-0007-9996-940X}{}

\author{José Martins}{Centro ALGORITMI / LASI, Universidade do Minho, Portugal}{jose.martins@dei.uminho.pt}{https://orcid.org/0000-0001-9380-7150}{Supported by FCT grant SFRH/BD/138660/2018}

\author{Sandro Pinto}{Centro ALGORITMI / LASI, Universidade do Minho, Portugal}{sandro.pinto@dei.uminho.pt}{https://orcid.org/0000-0003-4580-7484}{}

\authorrunning{Diogo Costa et al.} 

\Copyright{Diogo Costa, Gonçalo Moreira, Afonso Oliveira, Jose Martins, Sandro Pinto} 



\category{} 

\relatedversion{} 

\EventEditors{Patrick Meumeu Yomsi and Stefan Wildermann}
\EventNoEds{2}
\EventLongTitle{Sixth Workshop on Next Generation Real-Time Embedded Systems (NG-RES 2025)}
\EventShortTitle{NG-RES 2025}
\EventAcronym{NG-RES}
\EventYear{2025}
\EventDate{January 20, 2025}
\EventLocation{Barcelona, Spain}
\EventLogo{}
\SeriesVolume{128}
\ArticleNo{2}

\newcommand{\mypara}[1]{\vspace{7pt}\noindent{\textit{\textbf{#1}}}}

\begin{document}

\maketitle

\begin{abstract}

Modern embedded systems are evolving toward complex, heterogeneous architectures to accommodate increasingly demanding applications. Driven by industry SWAP-C (Size, Weight, Power, and Cost) constraints, this shift has led to the consolidation of multiple systems onto single hardware platforms. Static Partitioning Hypervisors (SPHs) offer a promising solution to partition hardware resources and provide spatial isolation between critical workloads. However, shared hardware resources like the Last-Level Cache (LLC) and system bus can introduce significant temporal interference between virtual machines (VMs), negatively impacting performance and predictability.
Over the past decade, academia and industry have focused on developing interference mitigation techniques, such as cache partitioning and memory bandwidth reservation. Configuring these techniques, however, is complex and time-consuming. Cache partitioning requires careful balancing of cache sections across VMs, while memory bandwidth reservation requires tuning bandwidth budgets and periods. With numerous possible configurations, testing all combinations is impractical and often leads to suboptimal configurations. Moreover, there is a gap in understanding how these techniques interact, as their combined use can result in compounded or conflicting effects on system performance.
Static analysis solutions that estimate worst-case execution times (WCET) and upper bounds on execution times provide some guidance for configuring interference mitigation techniques. While useful in identifying potential interference effects, these tools often fail to capture the full complexity of modern multi-core systems, as they typically focus on a limited set of shared resources and neglect other sources of contention, such as IOMMUs and interrupt controllers.
To address these challenges, we introduce SP-IMPact, an open-source framework designed to analyze and guide the configuration of interference mitigation techniques, through the deployment of diverse VM configurations and setups, and assessment of hardware-level contention (leveraging SPHs). 
It supports two mitigation techniques: (i) cache coloring and (ii) memory bandwidth reservation, while also evaluating the interactions between these techniques and their cumulative impact on system performance. By providing insights on real hardware platforms, SP-IMPact helps to optimize the configuration of these techniques in mixed-criticality systems, ensuring both performance and predictability.

\end{abstract}

\section{Introduction}
\label{sec:intro}

In recent decades, a significant trend toward digitization has revolutionized various industries including automotive, robotics, medical, and aerospace \cite{cerrolaza2020multi, LTZVisor2017, jailhouse2017}. This shift brought an exponential increase in system features, prompting high-end embedded platforms to evolve from basic designs. Past simple MCUs with single cores have given way to today's intricate and highly complex platforms \cite{costa2023}. The transition from single-core to multi-core architectures, accommodating multiple CPUs, and integrating diverse hardware accelerators like \ac{GPUs}, \ac{TPUs}, \ac{NPUs}, and \ac{FPGAs} \cite{gracioli2019, mancuso2013real, IAAI2024}, has fundamentally altered the landscape, resulting in highly heterogeneous designs.

Simultaneously, market demands for compact and efficient systems have driven the consolidation of multiple functionalities onto single hardware platforms to meet \ac{SWaP-C} constraints. This consolidation has led to the rise of \ac{MCSs} \cite{henzinger2006embedded}, where components with varying criticality levels coexist on the same platform. Virtualization technologies have been instrumental in enabling such consolidation, with hypervisor-based solutions—particularly static-partitioning hypervisors \cite{martins2023shedding, bao2020, jailhouse2017, hwang2008xen, klein2009sel4}—striking a balance between safety, security, and resource efficiency. These hypervisors allow for the deployment of diverse workloads within \ac{MCSs} while adhering to stringent industry safety standards, such as ISO 26262 \cite{ISO26262}.

Achieving robust system consolidation requires addressing critical challenges to ensure safety and security, particularly spatial and temporal isolation. Spatial isolation guarantees that architectural resources (e.g., CPUs and main memory) allocated to one system remain inaccessible to others. Temporal isolation ensures that the execution of one system’s workloads does not interfere with another's timing requirements. While static partitioning effectively addresses spatial isolation, temporal isolation remains a significant challenge due to contention on shared microarchitectural resources like the \ac{LLC}, main memory, and system bus. Such contention leads to increased execution times and reduced determinism \cite{abella2015wcet, cazorla2019probabilistic, cazorla2013proartis, Tamara2022, abella2015wcet}, making timing predictability particularly difficult for hard real-time systems.

Techniques such as cache partitioning \cite{Gracioli2015} and memory bandwidth reservation \cite{MemGuard} have emerged as promising solutions to mitigate temporal interference in \ac{MCSs}. Cache partitioning segments the \ac{LLC} into regions assigned to specific \ac{VMs}, while memory bandwidth reservation regulates the number of memory accesses within a given time frame. However, configuring these techniques effectively requires careful balancing of resources across \ac{VMs} and fine-tuning parameters (e.g., define cache regions and/or memory budgets and periods). This process is complex, time-consuming, and impractical for real-world \ac{MCSs}, often leading to suboptimal configurations.
Static analysis tools \cite{abella2015wcet, arora2022bus} have been explored to address these challenges, offering a means to understand interference impacts in \ac{MCSs} and guide the configuration of mitigation techniques. By estimating \ac{WCET} and quantifying interference effects, these tools provide a foundation for informed decision-making. However, existing static analysis solutions often focus on specific shared resources, such as the \ac{LLC}, overlooking other shared hardware resources (e.g., IOMMUs and interrupt controllers). 

To address these limitations, we introduce SP-IMPact, an open-source framework \footnote{https://gitlab.com/ESRGv3/sp-impact} designed to analyze and support the configuration of interference mitigation techniques. SP-IMPact enables a comprehensive understanding of the impact of shared hardware resources on real platforms by considering all potential sources of contention - filling critical gaps left by currently available solutions. Furthermore, it allows the deployment of diverse configurations of cache coloring and memory bandwidth reservation to evaluate their effects on the workloads of different \ac{VMs}. The insights gained through SP-IMPact can guide the optimization of these techniques, easing their usage by industry due to the framework's workload-agnostic design, which supports various operating systems and workloads. For academia, SP-IMPact provides a versatile launchpad for deploying and testing new interference mitigation techniques. While this paper leverages the Bao hypervisor \cite{bao2020} as a use case, SP-IMPact is agnostic to the underlying hypervisor and can be extended to support other static partitioning hypervisors.

\section{Background}


The consolidation of \ac{MCSs} introduces a well-known challenge: interference between co-existing \ac{VMs}, which can degrade performance and disrupt real-time guarantees \cite{gracioli2019, bao2020, kloda2019deterministic, yun2012memory, Yun2014, MemGuard, Bechtel2019, ungerer2010merasa, dasari2013identifying, lofwenmark2016understanding, kotaba2013multicore}. This interference typically arises from contention over shared micro-architectural resources, such as the \ac{LLC}, main memory, and the system bus, leading to increased execution time and lack of determinism. 
To mitigate these issues, techniques such as cache partitioning \cite{kloda2019deterministic, Modica2018, Kim2017} and memory bandwidth reservation \cite{MemGuard, yun2012memory, Modica2018, Farshchi2020}, among others solutions targeting I/O and interrupts regulation \cite{Zini2022, gordon2012eli, costa2022} , have been proposed.

\begin{figure}[t] 
    \centering 
    \subfloat[\centering Cache Coloring Toy Example]{{\includegraphics[width=.49\linewidth]{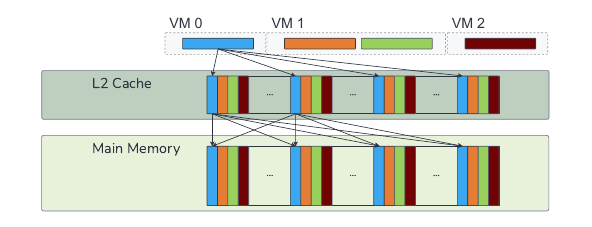} }}
    \subfloat[\centering Memory Bandwidth Reservation Toy Example]{{\includegraphics[width=.49\linewidth]{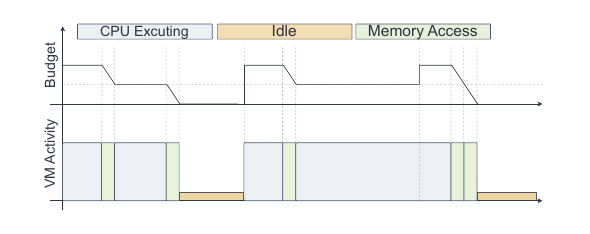} }} 
    \caption{Illustrative examples of cache coloring and memory bandwidth reservation mechanisms.} 
    \label{fig:interf_mitigation_techniques_soa} 
\end{figure}

\mypara{Cache Partitioning:} 
Cache partitioning techniques enable the selective allocation of cache regions to specific workloads, thereby reducing cache contention and improving predictability. One of the most used techniques for cache partitioning is cache coloring, which divides the \ac{LLC} into distinct regions by assigning specific "colors" to individual workloads, where each "color" corresponds to different cache sets, which helps to control cache access patterns and minimize interference. Cache coloring is commonly used to assign dedicated portions of the cache to individual VMs, effectively limiting \ac{LLC} contention, as depicted in Figure \ref{fig:interf_mitigation_techniques_soa} (a).

\mypara{Memory Bandwidth Reservation:} Memory bandwidth reservation techniques regulate the memory access rate to reduce interference and ensure temporal isolation. MemGuard \cite{MemGuard}, for example, limits the memory access rate per CPU by allocating specific portions of memory bandwidth, or "budgets," over defined periods, as depicted in Figure \ref{fig:interf_mitigation_techniques_soa} (b). This prevents any workload from monopolizing memory, ensuring fair resource distribution and reducing delays, thus improving predictable performance and minimizing interference between workloads. 

\section{System Overview}

\begin{figure}[t]
    \centering
    \includegraphics[width=1.0\linewidth]{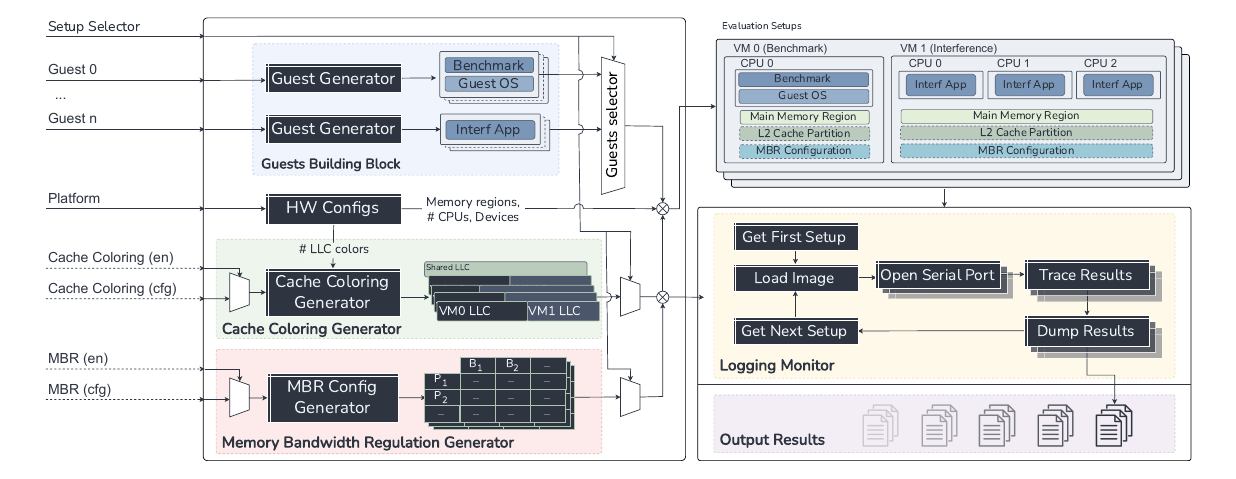}
    \caption{SP-IMPact System Overview.}
    \label{fig:BaoRTI_SystemOverview}
\end{figure}

In this paper, we introduce SP-IMPact, a framework developed to evaluate and benchmark the performance of \ac{MCSs}, with a focus on measuring interference and assessing the effectiveness and interaction of interference mitigation techniques. 
Using a configuration file, users can specify the platform, guest definitions, and test setups. As shown in Figure \ref{fig:BaoRTI_SystemOverview}, the framework leverages three key components: (i) the \textbf{Guest Generator}, which build guests; (ii) the \textbf{Cache Coloring Generator}, which defines cache partitioning configurations; and the (iii) \textbf{Memory Bandwidth Regulation Generator}, which generates different configurations of the mechanism. Together, these components ensure precise and consistent performance evaluations. SP-IMPact also includes a \textbf{Logging Monitor} for collecting run-time data and an \textbf{Output Results} module to handle the gathered information for further analysis.

\mypara{Guests Generator:} In the context of evaluating system performance, the framework provides support for constructing different types of guests, each tailored for specific benchmarking or interference generation tasks. For the sake of simplicity, this discussion focuses on two primary guest types: a Linux guest and a baremetal guest. However, it should be noted that the framework can be extended to support more and varied guest types as needed.

\begin{enumerate}
    \item \textbf{Linux Benchmark}: Designed to simplify the deployment of various Linux-based workloads, enabling the evaluation of system behavior across diverse scenarios.

    \item \textbf{Contention Engine}: A baremetal guest tailored to create memory and hardware resource pressure, targeting the \ac{LLC} and main memory. Key parameters include CPU count, workload sizes, and operation types (reads, writes, or both).
\end{enumerate}

To formalize these configuration options for the baremetal guest, let \( \textbf{\textit{M}} \) represent the total number of CPUs available, \( \textbf{\textit{L}} \) denote the set of possible cache line sizes, \( \textbf{\textit{W}} \) signify the set of workload sizes, and \( \textbf{\textit{O}} \) define the set of operation types, where \( \textbf{\textit{O}} = \{ \text{read}, \text{write}, \text{read/write} \} \). The configuration space for the baremetal guest can thus be expressed as:
\begin{equation}
\label{eq:baremetal_guests}
    G_{\text{baremetal}} = \{ (m, l, w, o) \mid m \in M, l \in L, w \in W, o \in O \}
\end{equation}
where \( G_{\text{baremetal}} \) represents the set of all possible configurations for the baremetal guest.
To represent the complete configuration space for all guests, including both baremetal and Linux guests, let \( G \) denote the overall set of guest configurations, which is equal to \( G_{\text{baremetal}} \cup G_{\text{linux}} \),
where \( G_{\text{linux}} \) represents the set of configurations for the Linux guest.


\mypara{Cache Coloring Generator:} In \ac{MCSs}, cache partitioning is crucial for reducing \ac{LLC} interference and ensuring predictable performance. This process, tipicaly based on cache coloring, aims to divide cache sets into non-overlapping regions for each VM. To assess the impact of different cache partitions, the BaoRTI framework supports the generation of distinct cache color configurations based on the following parameters:

\begin{enumerate}
    \item The total number of cache sets \( \textit{\textbf{S}} \), which corresponds to the bit length of the bitmap used to define the color assignments (i.e., each cache set index is represented as a bit in the range \([0, S - 1]\));
    \item The number of VMs, \( \textit{\textbf{N}} \), to which distinct cache partitions will be assigned.
\end{enumerate}

The objective of the function is to generate unique configurations of bit masks, dividing the bit range \([0, S - 1]\) into \( \textit{\textbf{N}} \) distinct non-overlapping sections. Each section represents a cache partition that can be assigned as a color to the VMs. Given \( \textit{\textbf{S}} \) cache sets and \( \textit{\textbf{N}} \) VMs, the function produces a unique configuration of non-overlapping bit masks for each VM. 
To facilitate this process, we denote \( \textit{\textbf{C}} \) as the set of all possible \((N - 1)\)-combinations of bit positions within the range \([0, S - 1]\). Each combination in \( \textit{\textbf{C}} \) represents potential VM coloring configurations and can be formally defined as:
\begin{equation}
    \textit{\textbf{C}} = \{ (b_1, b_2, \dots, b_{N-1}) \mid 0 \leq b_1 < b_2 < \dots < b_{N-1} < S \}
\end{equation}
where each \( b_i \) denotes a bit position that separates the partitions for each VM. 


To generate the bit masks for each VM, we begin by initializing the starting bit \( s_0 \) to \( 0 \). For each VM \( i \), where \( i \) ranges from \( 0 \) to \( N - 1 \), we define the end bit \( e_i \) based on the boundary positions: if \( i < N - 1 \), \( e_i \) is set to \( b_i \), whereas for the last VM (\( i = N - 1 \)), \( e_i \) is assigned the total number of cache sets \( S \). With these boundaries established, we compute the bit mask \( M_i \) for each VM using the formula:
\begin{equation}
    M_i = ((1 \ll (e_i - s_i)) - 1) \ll s_i
\end{equation}
where "\( \ll \)" represents a left bit shift operation. This formula creates a mask with \( (e_i - s_i) \) bits set to 1, aligned to begin at the position defined by \( s_i \). After calculating the bit mask, we update the starting bit \( s_i \) for the next VM by setting it to the current end bit \( e_i \). This iterative process continues until all masks for the VMs are generated, ensuring that each VM receives a unique configuration of non-overlapping cache partitions.
After generating a list of bit masks for each VM in the current combination, this configuration is added to a result set \( \text{colors\_assignments} \) if it does not already exist in the set. This ensures that all configurations in that list are unique.

\mypara{Memory Bandwidth Regulation Generator:} In real-time systems, generating distinct memory bandwidth configurations for \ac{VMs} is essential for ensuring predictable performance and efficient resource utilization. This process, known as memory bandwidth reservation, focuses on creating unique combinations of budget and sampling period for each VM. To evaluate the effects of various bandwidth configurations, the BaoRTI framework supports the generation of distinct MBR configurations based on the following parameters:

\begin{enumerate}
    \item A list of budgets \( \textit{\textbf{B}} \) available for reservation, where each budget specifies the maximum amount of bandwidth allocated to a VM.
    \item A list of sampling periods \( \textit{\textbf{P}} \), which define the time intervals at which the allocated bandwidth should be monitored.
\end{enumerate}

The objective of the memory bandwidth reservation assignment generation is to produce all the possible combinations of budget and period assignments for each guest. Given the set of all guests, \( \textbf{\textit{G}} \), where each guest, \( \textbf{\textit{g}} \),  is associated with a set of budgets, \( B_g \), and a set of periods, \( P_g \), the configuration for each guest can be expressed as:
\begin{equation}
    MBR_g = \{ (B, P) \mid B \in B_g, P \in P_g \}
\end{equation}
where \( MBR_g \) represents the set of all combinations of memory bandwidth reservation configurations for the guest \( g \). 
Let \( G \) be the total number of guests, \( B \) be the maximum number of budgets across guests, and \( P \) be the maximum number of sampling periods across guests. The time complexity for generating the budget-period combinations for each guest is \( O(B \times P) \). Since there are \( G \) guests, the overall complexity for processing all guests and generating their combinations can be expressed as \(  O\left(G \times B \times P\right) \).



SP-IMPact features a results logging system designed to capture essential performance and behavioral metrics from the target platform during test execution. The framework collects data from multiple serial ports, each mapped to a specific \textbackslash{}ac\{VM\}, ensuring comprehensive monitoring across the system.
The captured metrics include execution time and key micro-architectural events, such as ac{LLC} misses, memory access counts, and cycles spent on the system bus. These metrics are vital for evaluating the impact of shared hardware resources—like the \ac{LLC} and memory controllers—on workload performance and predictability.
This versatile design enables SP-IMPact to support a wide array of benchmarks and metrics tailored to diverse interference scenarios. By correlating data across multiple \ac{VMs} and configurations, SP-IMPact provides the granularity required to assess the effectiveness of interference mitigation techniques and optimize their configurations.

\section{Evaluation}
\subsection{Evaluation Setup}

\mypara{Hardware Platform.} The experiments were conducted on a Xilinx ZCU104 evaluation board equipped with a Zynq Ultrascale+ ZU7EV SoC. This platform includes a quad-core Arm Cortex-A53 processor, operating at 1.2 GHz. While the SoC supports up to 16 distinct cache colors for cache coloring, the Bao hypervisor constrains this to 8 colors to avoid partitioning the L1 cache. Each core has a dedicated 32 KiB L1 instruction and data cache, along with a unified 1 MiB L2 cache. Additionally, the board is equipped with an Arm Performance Monitoring Unit (PMU), which was leveraged to collect microarchitectural events (such as cache misses and system bus accesses) and profile the benchmark.

\mypara{Workloads.} For our evaluation, we leveraged the MiBench Automotive and Industrial Control System (AICS) \cite{guthaus2001mibench} Suite within the critical VM. This subset includes three memory-intensive benchmarks: 
\textit{qsort}, \textit{susan-c}, and \textit{susan-e}. 
To generate interference at the memory hierarchy, we deployed a baremetal application that continuously performs read or write operations on a buffer with different sizes. Specifically, buffer sizes include 32 KiB (100\% of the L1 cache), 512 KiB (50\% of the L2 cache), 1 MiB (100\% of the L2 cache), 1.5 MiB (150\% of the L2 cache), 2 MiB (200\% of the L2 cache), and 4 MiB (400\% of the L2 cache).

\mypara{Setups.} According to Equation \ref{eq:baremetal_guests}, we consider the following parameters: \( L = 1 \)~ (using only the cache line size matching the cache line size of the target hardware platform), \( C = 1 \) (using only one CPU configuration, which assigns 3 CPUs to the baremetal VM), \( W = 6 \) (the total number of workloads used), and \( O = 2 \) (representing both read and write operations). This results in a total of 12 variations of the baremetal guest.
Additionally, for cache coloring, since there are 2 VMs (\( N = 2 \)) and 8 possible cache sets (\( S = 8 \)), there are 8 unique configurations of cache coloring. However, we excluded scenarios in which the Linux VM would be allocated only a single cache color, as such configurations would not provide meaningful performance benefits. 
Thus, combining these factors results in a total of 84 setups to be tested.
For simplicity, we will not consider the configuration of MBR, as introducing it would significantly increase the total number of setups in this evaluation section. 

\mypara{Setup Naming Convention.} Setups are named \texttt{solo} or \texttt{interf\_<access>\_<buffer\_size>}. The \texttt{solo} setup serves as the baseline, where a Linux VM runs the MiBench benchmarks without interference. In \texttt{interf\_<access>\_<buffer size>} setups, an additional workload creates cache contention, with \texttt{access} specifying \texttt{read} or \texttt{write} interference type and \texttt{buffer\_size} indicating the buffer size used. Cache coloring setups add the suffix \texttt{<cc\_num-colors>}, where \texttt{num-colors} denotes the cache colors allocated to the critical VM.

\subsection{Interference Impact on Multi-core Platforms}

\begin{figure}[t]
    \centering
    \includegraphics[width=0.99\linewidth]{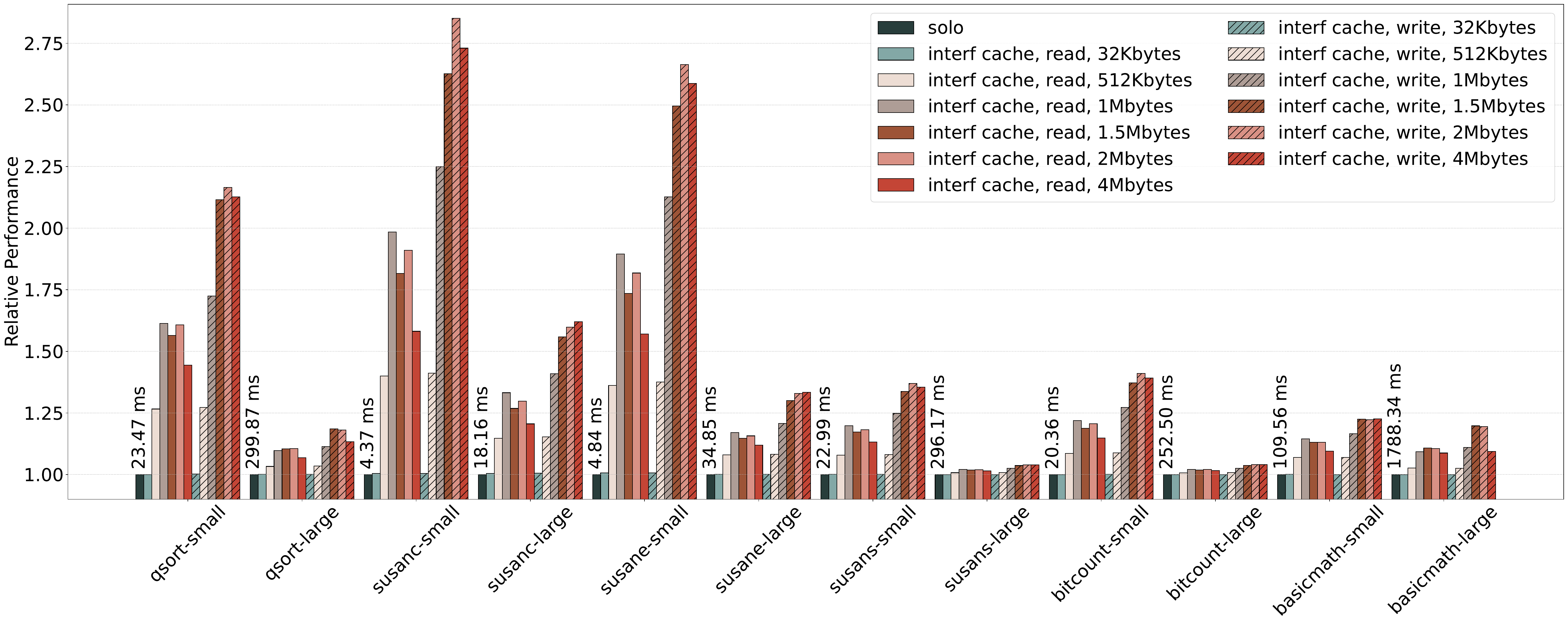}
    \caption{Performance overheads of MiBench automotive benchmark with different workloads}
    \label{fig:mibench_benhmark}
\end{figure}

\begin{figure}[t] 
    \centering 
    \subfloat[\centering Interference Impact on LLC]{{\includegraphics[width=.49\linewidth]{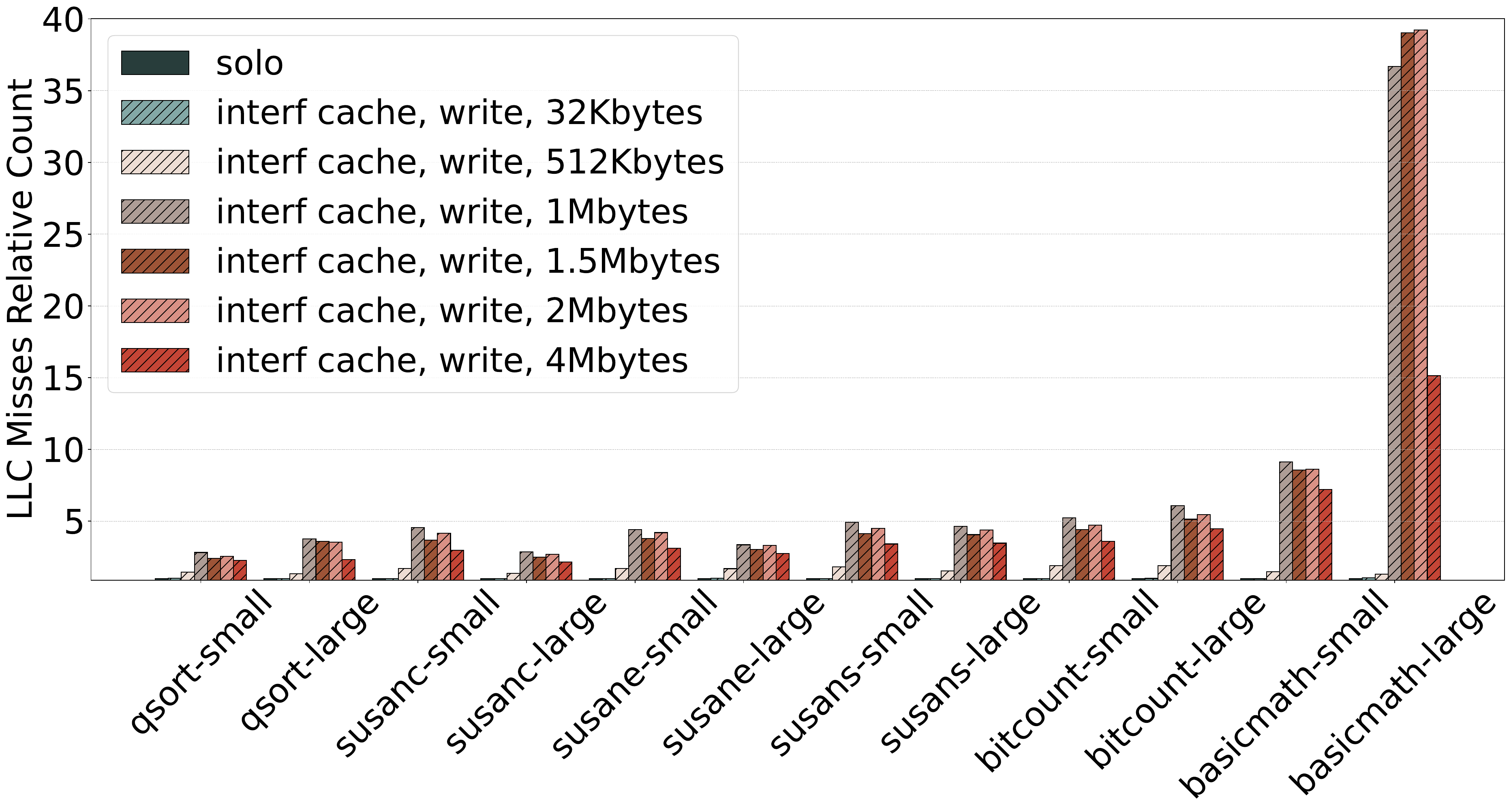} }}
    \subfloat[\centering Interference Impact on system bus]{{\includegraphics[width=.49\linewidth]{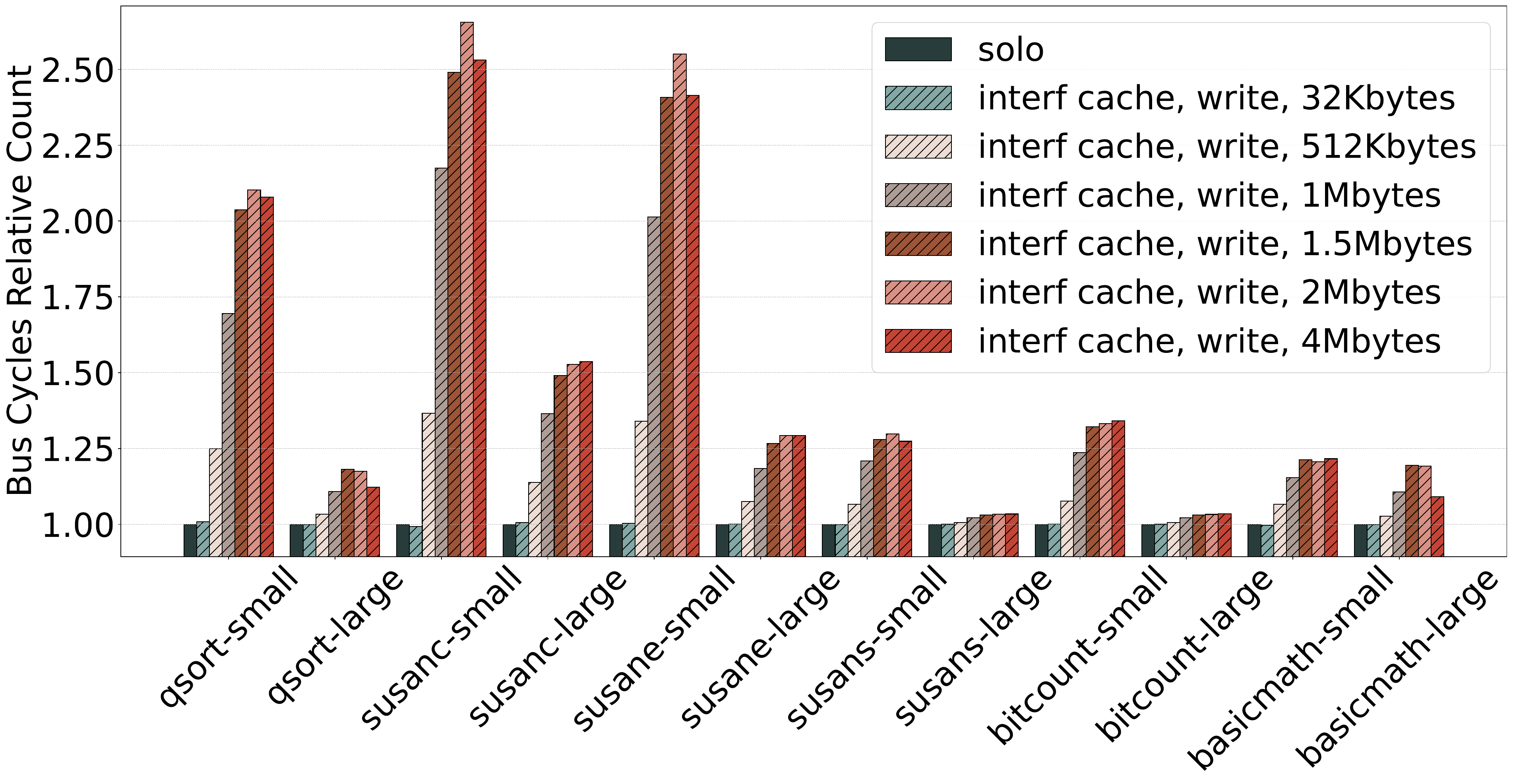} }}
    \caption{Collected PMU events from MiBench benchmark} 
    \label{fig:interf_impact_pmu}
    \label{fig:interf_mitigation_techniques} 
\end{figure}


Empirical results presented in Figure \ref{fig:mibench_benhmark} indicate that contenetion on shared hardware resources can severely hamper the performance of memory-intensive benchmarks such as \textit{qsort-small}, \textit{susan-c-small}, and \textit{susan-e-small}. The results confirm the theoretical expectations of how the interference buffer size influences resource contention, providing valuable insight into the SP-IMPact framework’s role in identifying and quantifying such issues. This framework proves essential in assessing how system configurations can exacerbate or mitigate performance bottlenecks in multi-core platforms. The observed interference patterns, where larger buffer sizes lead to increased contention for shared resources, underscore the importance of understanding system-level interactions in \ac{MCSs}.

\begin{figure}[t]
    \centering
    \includegraphics[width=.99\linewidth]{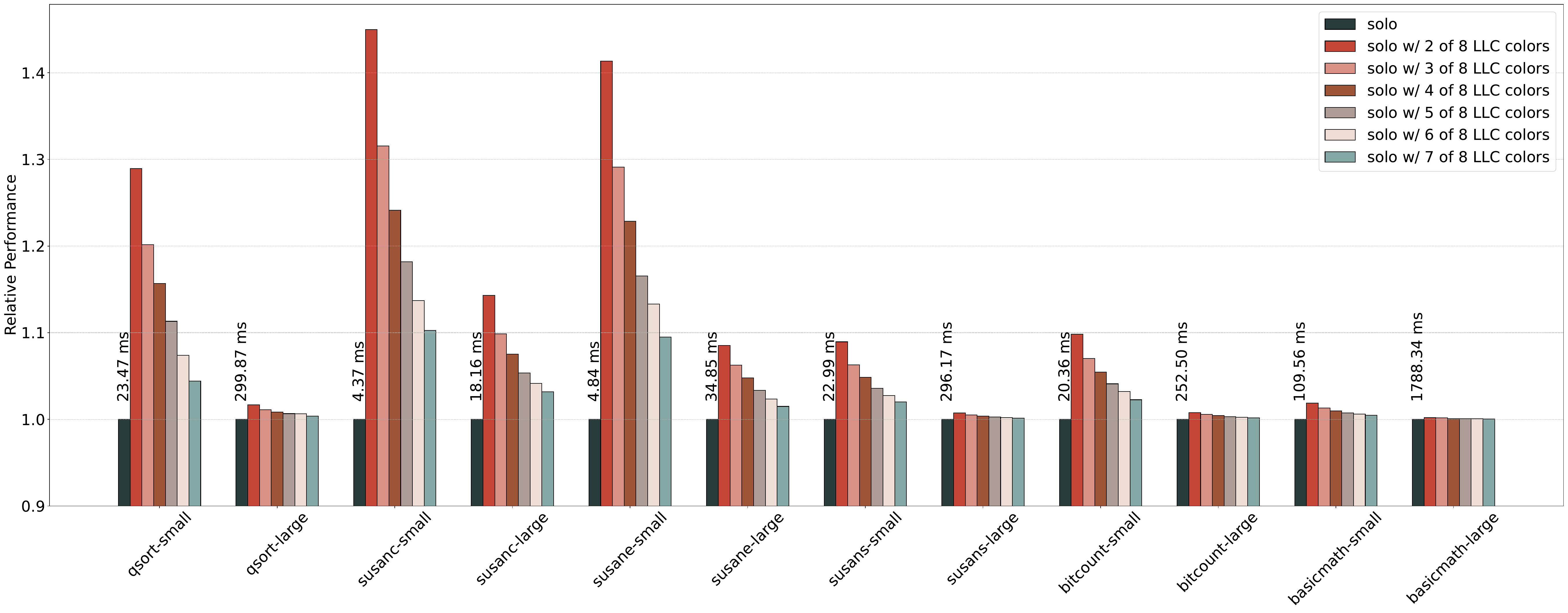}
    \caption{Cache coloring configuration impact on MiBench Benchmark}
    \label{fig:enter-label}
\end{figure}

\begin{figure}[t]
    \centering
    \subfloat[\centering Cache Coloring on 1MiB interference scenario]{{\includegraphics[width=.99\linewidth]{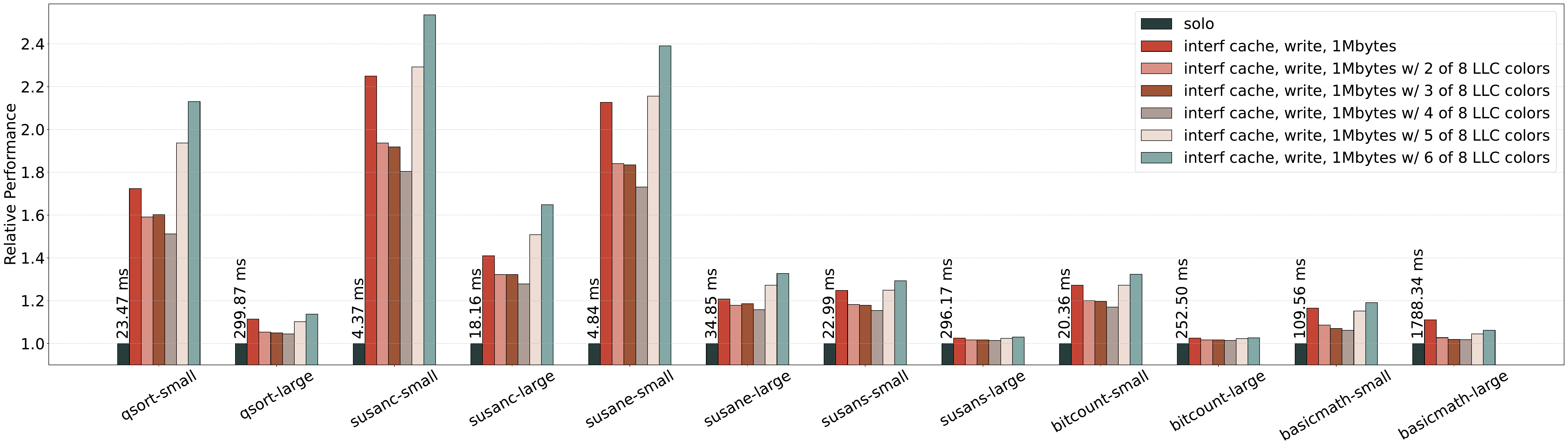} }} \\ \vspace{.5cm}
    \subfloat[\centering Cache Coloring on 2MiB interference scenario]{{\includegraphics[width=.99\linewidth]{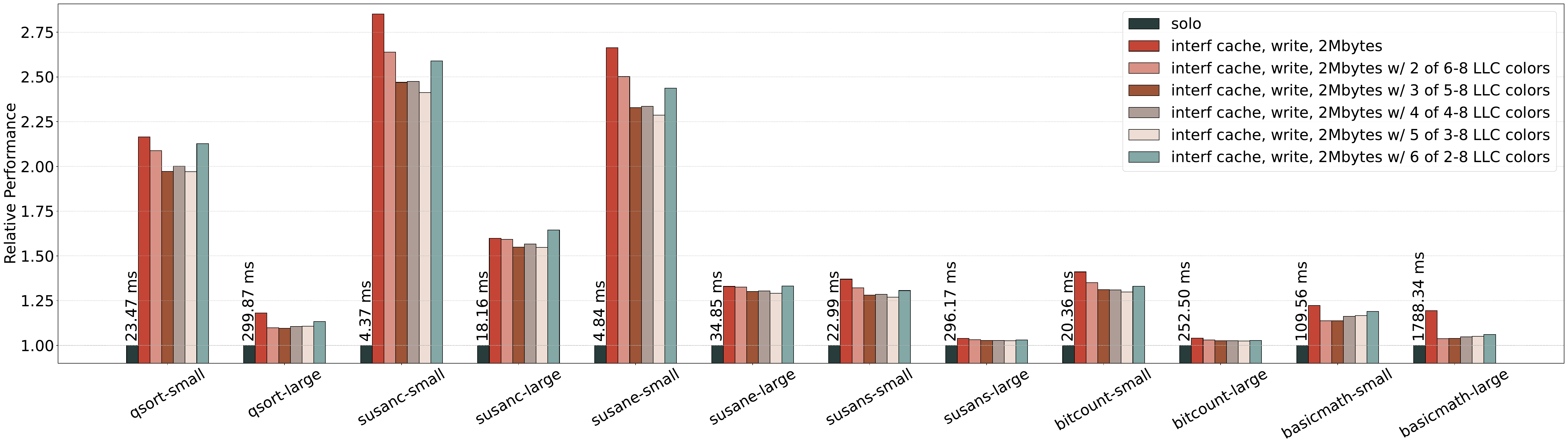} }} 
    \caption{Cache Coloring interference mitigation on MiBench Benchmark}
    \label{fig:cc_interf_mit_mibench_pmu}
\end{figure}

While the the empirical resutls where theoretically expected, the empirical evidence reinforces the critical need for tools like SP-IMPact to understand the impact of consolidating different workloads on top of the same hardware platform. Not only does the framework help identify these issues, but it also enables developers to quantify the effects of interference under different configurations, a key insight to drive the deployment of \ac{MCSs}.
By running different workloads with different configurations, developers can collect key performance metrics, such as execution time, cache misses, and bus cycles, which are essential for understanding the severity of the interference. These metrics provide a comprehensive view of how shared resources (e.g. the \ac{LLC} and the system bus) impact overall system performance. For example, as depicted in Figure \ref{fig:interf_impact_pmu}(a), increasing the buffer size leads to a notable rise in cache misses, which in turn increases the execution time. Specifically, in the \texttt{interf\_write\_1MiB} scenario, the execution time for the \textit{susanc-small} benchmark increases from 4.37 ms to 9.83 ms, demonstrating the growing impact of interference as the buffer size increases.

The role of the SP-IMPact framework in identifying these performance impacts is critical, as it helps pinpoint where interference is most pronounced. Once performance metrics are gathered, the framework allows for in-depth analysis to identify the root causes of performance degradation. For instance, as the interference buffer size grows, portions of the L2 cache become occupied, leading to cache contention and cache evictions. These evictions result in increased memory access time, contributing to further performance slowdowns. The underlying mechanism driving this issue is the competition for cache lines, which causes more frequent evictions and delays in data retrieval. This phenomenon is compounded by the finite size of the cache, which limits the amount of data that can be stored and retrieved quickly. Additionally, Figure \ref{fig:interf_impact_pmu}(b) shows that increasing the buffer size also introduces contention on the system bus, further exacerbating the performance overhead. As workloads compete for access to shared bus resources, the time spent transferring data between the CPU and memory increases, leading to a marked decline in overall system efficiency. These findings underscore the importance of managing resource contention in multi-core environments, where shared hardware resources are increasingly stressed by demanding workloads.

\subsection{Interference Mitigation Techniques}

\mypara{Cache Coloring Overhead.} Empirical results shows that cache coloring, even in a single-VM environment without interference, can introduce variations in benchmark performance depending on the number of cache colors available for the benchmark’s use. This effect is most pronounced in memory-intensive benchmarks, such as \textit{qsort-small}, \textit{susan-c-small}, and \textit{susan-e-small}, where reduced cache availability leads to significant slowdowns due to increased cache misses. 
With only two of the eight cache colors available, the execution time of benchmarks like \textit{susanc-small} and \textit{susane-small} increases by 1.45x and 1.41x, respectively, due to reduced cache allocation. 
As the number of colors increases, performance gradually approaches the baseline. With five cache colors available, benchmarks generally perform closer to their solo execution times. For example, \textit{susanc-small} and \textit{susane-small} improve to slowdowns of 1.18x and 1.17x, respectively. Benchmarks with lower memory intensity, such as \textit{qsort-large}, \textit{basicmath-small}, and \textit{basicmath-large}, show minimal to no performance degradation across various cache coloring scenarios. With only two colors, \textit{basicmath-large} shows no measurable slowdown across all coloring configurations. Similarly, \textit{qsort-large} and \textit{basicmath-small} maintain near-baseline performance, with minimal slowdowns of 1.02x.

\mypara{Interference Mitigation.} 
The SP-IMPact framework plays a key role in assisting developers with mitigating memory contention issues during the development of \ac{MCSs}. After identifying bottlenecks caused by shared hardware resources, developers can leverage the framework to simulate different scenarios and adjust system configurations accordingly. For example, cache coloring can be leveraged to minimize interference. Figures \ref{fig:cc_interf_mit_mibench_pmu}(a) and \ref{fig:cc_interf_mit_mibench_pmu}(b) show the impact of different cache coloring configurations on memory-intensive benchmarks, such as \textit{qsort-small} and \textit{susanc-small}, when consolidated with the interference baremetal VM (e.g, running the \texttt{interf\_write\_1MiB} and the \texttt{interf\_write\_2MiB} scenarios). Applying 2 cache colors reduces execution time overhead from 1.72x and 2.25x (no coloring) to 1.59x and 1.94x, respectively. Further improvements are observed with 4 cache colors, reducing interference to 1.51x and 1.80x for \textit{qsort-small} and \textit{susanc-small}. While cache coloring is effective for memory-intensive workloads, developers should consider diminishing returns beyond 4 colors, where performance gains decrease, and system-level contention (especially on the bus) may increase. The SP-IMPact framework helps identify these diminishing returns, allowing developers to select the most optimal configuration. For less memory-intensive benchmarks like \textit{qsort-large} and \textit{basicmath-large}, cache coloring has minimal impact, enabling developers to focus on other optimization techniques for such workloads.

\section{Discussion and Future Directions}

In this section, we discuss some of the open issues and potential research directions to understand and improve the impact of interference in multi-core platforms.

\mypara{Workload Interference Analysis in Mixed-Criticality Systems.} \ac{MCSs} face significant challenges when consolidating workloads with varying criticality levels and timing requirements on the same hardware platform. As workloads compete for shared resources such as caches, memory buses, and system interconnects, predicting interactions and maintaining reliable performance for critical tasks remains a complex problem. While the framework presented in this work enables profiling and quantifying interference effects under diverse scenarios, further research is needed to explore how workload characteristics - such as memory access patterns and computation intensity (e.g., memory access rate - can be modeled more accurately. Moreover, one important limitation of the current evaluation is that it primarily focuses on interference effects in terms of LLC misses and bus cycles; 
while the SP-IMPact framework allows the analysis of the contention in these components, the lack of state-of-the-art benchmarks to evaluate them limits their inclusion in this study.

\mypara{Interference Mitigation Techniques.} Configuring interference mitigation mechanisms, such as cache coloring, presents its own set of challenges. Each possible configuration (e.g. the number of cache colors or memory bandwidth regulation configuration) can produce different impacts on performance and contention levels. Selecting the optimal configuration requires an understanding of both the workload's memory demands and the system's architectural characteristics. The framework aids in this process by providing in-depth evaluations of various interference mitigation techniques configurations and their impact on interference on multi-core platforms. Additionally, exploring the interaction between the proposed framework and high-performance hardware features, such as quality-of-service (QoS) mechanisms that control on-chip and DRAM traffic, would be valuable, as presented in \cite{serrano2021leveraging}.

\mypara{Future Work.} Building on the insights from this study, several extensions to the current framework are planned: an immediate enhancement of SP-IMPact involves the development of a hypervisor-level performance monitor to enable VM profiling without requiring guest instrumentation.
Currently, the framework simplifies the generation of Linux-based benchmarks and baremetal VMs; next-steps focus on extending this capability to other OSes (e.g., FreeRTOS and Zephyr), enabling a comprehensive interference analysis and mitigation evaluations across a wider range of workloads and system configurations.
In the long term, we aim to integrate AI-driven techniques for adaptive interference management, which may enable the optimization of interference patterns, allowing the simulation of worst-case scenarios and providing more accurate performance assessments under challenging conditions.

\section{Related Work}

Interference analysis in multi-core systems has been approached using two primary frameworks: (i) generic task models and (ii) phased execution models. Each approach has its strengths but also limitations when it comes to capturing the complexities of modern high-complexity \ac{MCSs}, especially those that include hypervisors. In the following, we provide a brief overview of these two categories and highlight the key research efforts within each.

\mypara{Generic Task Models.} Generic task models provide abstractions for task behaviors on multi-core systems by focusing on resource usage patterns such as memory access, computation, and synchronization. These models typically focus on quantifying contention between tasks based on broad assumptions and often omit platform-specific characteristics. Generic task models can be divided into two main categories: 
(i) memory bus contention and \cite{chattopadhyay2010modeling, rosen2007bus, schranzhofer2010timing, kelter2011bus, jacobs2016framework, davis2018extensible},
(ii) main memory contention \cite{yun2015parallelism, kim2014bounding, hassan2018bounding}.

\mypara{Phased Execution Models.} While generic task models provide broad abstractions, they are limited in their ability to model complex, dynamic interference patterns that arise in multi-core systems. This limitation led to the development of phased execution models, which break down task execution into distinct phases. Phased execution models offer more detailed representations of how tasks interact with shared resources during different execution phases. To address the issues left by generic task models, phased execution models are divided in: 
(i) offline scheduling-based approaches \cite{senoussaoui2022contention, pagetti2018automated, becker2016contention}, 
(ii) shared resource contention-based approaches \cite{maia2017schedulability, casini2020holistic, arora2022bus}, and 
(iii) memory-centric scheduling-based approaches \cite{yao2015global, schwaricke2020fixed}.

\mypara{Limitations of Existing Approaches.} While generic task and phased execution models help understand some aspects of interference, they fall short in high-complexity \ac{MCSs}, especially those with hypervisors. These models focus on limited contention sources, like LLC and system buses, but omit others such as IOMMUs or interrupt controllers, which are crucial in real hardware. Moreover, they overlook the combined effects of multiple mitigation techniques.
SP-IMPact fills these gaps by enabling the assessment of interference in hypervisor-based systems and evaluating the effectiveness and interactions of interference mitigation techniques.
Unlike analytical models, SP-IMPact simplifies configuration by allowing real-time experimentation on actual hardware, making it easier to identify bottlenecks and test configurations in a flexible way.
The framework was tested with Bao hypervisor and currently supports its configuration interface to define VMs' configurations and hardware partitioning, but it can be extended to support other hypervisors in the future.

\section{Conclusion}

In this paper, we propose the design, implementation, and evaluation of SP-IMPact, a framework for analyzing the impact of multi-core contention, and the impact of interference mitigation techniques. This framework facilitates the automated deployment, configuration, and data collection of multiple setups to quantify platform-level contention and evaluate the impact of interference mitigation techniques, such as cache coloring. Using the Zynq Ultrascale+ platform, our evaluation demonstrated how the framework enables precise analysis of interference effects under various workload configurations, providing critical insights for deploying consolidated multi-core systems. We believe that this framework lays a solid foundation for future extensions, including AI-driven interference management, expanded workload patterns, and support for additional platforms.

\bibliography{oasics-v2021-sample-article}

\end{document}